\documentclass[prl,twocolumn,superscriptaddress,longbibliography,notitlepage]{revtex4-1}
\usepackage{amsmath,amssymb,amsthm,amsfonts,mathrsfs}
\usepackage{bm}
\usepackage{textcomp}
\usepackage{graphicx}
\usepackage{color}
\usepackage{tikz,pgf}%
\usetikzlibrary{shadings}
\usepackage{dsfont}
\usepackage{fontenc}
\usepackage{amsfonts}
\usepackage{booktabs}
\usepackage[colorlinks=true, letterpaper=true, pdfstartview=FitV, linkcolor=blue, citecolor=blue, urlcolor=blue]{hyperref}

\begin{document}

\title{Universal Approach to Magnetic Second-Order Topological Insulator}

\author{Cong Chen}
\thanks{C. Chen and Z. Song contributed equally to this work.}
\affiliation{Key Laboratory of Micro-nano Measurement-Manipulation and Physics (Ministry of Education), School of Physics, Beihang University, Beijing 100191, China}
\affiliation{Research Laboratory for Quantum Materials, Singapore University of Technology and Design, Singapore 487372, Singapore}

\author{Zhida Song}
\thanks{C. Chen and Z. Song contributed equally to this work.}
\affiliation{Department of Physics, Princeton University, Princeton, New Jersey 08544, USA}

\author{Jian-Zhou Zhao}
\affiliation{Sichuan Co-Innovation Center for New Energetic Materials, Southwest University of Science and Technology, Mianyang 621010, China}
\affiliation{Research Laboratory for Quantum Materials, Singapore University of Technology and Design, Singapore 487372, Singapore}

\author{Ziyu Chen}
\affiliation{Key Laboratory of Micro-nano Measurement-Manipulation and Physics (Ministry of Education), School of Physics, Beihang University, Beijing 100191, China}

\author{Zhi-Ming Yu}
\affiliation{Key Lab of Advanced Optoelectronic Quantum Architecture and Measurement (MOE), Beijing Key Lab of Nanophotonics and Ultrafine Optoelectronic Systems, and School of Physics, Beijing Institute of Technology, Beijing 100081, China}
\affiliation{Research Laboratory for Quantum Materials, Singapore University of Technology and Design, Singapore 487372, Singapore}

\author{Xian-Lei Sheng}
\email{xlsheng@buaa.edu.cn}
\affiliation{Key Laboratory of Micro-nano Measurement-Manipulation and Physics (Ministry of Education), School of Physics, Beihang University, Beijing 100191, China}
\affiliation{Research Laboratory for Quantum Materials, Singapore University of Technology and Design, Singapore 487372, Singapore}

\author{Shengyuan A. Yang}
\affiliation{Research Laboratory for Quantum Materials, Singapore University of Technology and Design, Singapore 487372, Singapore}

\begin{abstract}
We propose a universal practical approach to realize magnetic second-order topological insulator (SOTI) materials, based on properly breaking the time reversal symmetry in conventional (first-order) topological insulators. The approach works for both three dimensions (3D) and two dimensions (2D), and is particularly suitable for 2D, where it can be achieved by coupling a quantum spin Hall insulator with a magnetic substrate. Using first-principles calculations, we predict bismuthene on EuO(111) surface as the first realistic system for a 2D magnetic SOTI. We explicitly demonstrate the existence of the protected corner states. Benefited from the large spin-orbit coupling and sizable magnetic proximity effect, these corner states are located in a boundary gap $\sim 83$ meV, hence can be readily probed in experiment. By controlling the magnetic phase transition, a topological phase transition between a first-order TI and a SOTI can be simultaneously achieved in the system. The effect of symmetry breaking, the connection with filling anomaly, and the experimental detection are discussed.

\end{abstract}

\maketitle

The discovery of topological insulators (TIs) has ignited one of the most active fields in physics research~\cite{Hasan_RMP,Qi_RMP,Bansil_RMP,ShunQingShen_TI}. Recently, a variant of TIs,
the second-order topological insulator (SOTI), has been proposed and attracted great interest~\cite{ZhangFan_PRL2013,Hughes2017,Langbehn2017,SongZD2017,Hughes2017b,Schindler2018SA, Ezawa2018}. While a conventional (first-order) TI in $d$ dimensions has protected gapless states on its $(d-1)$-dimensional boundary, a SOTI features protected gapless states on its $(d-2)$-dimensional boundary.
For a three-dimensional (3D) SOTI, such topological boundary states appear at certain 1D hinges between gapped crystal faces; and for a 2D SOTI, they are localized at 0D corners of the sample.

Besides the conceptual breakthrough, the advance of the field must require suitable material platforms to realize such topological states. For conventional TIs, we already have an inventory of experimentally verified materials, e.g., the Bi$_2$Se$_3$ family~\cite{HJZhang_Bi2Se3,Hasan_Bi2Se3,Chen178} for 3D TIs; and HgTe/CdTe quantum wells~\cite{Bernevig_HgTe,SCZhang_HgTe}, InAs/GaSb quantum wells~\cite{LiuCX_PRB2008,DuRR_PRL2011}, bismuthene~\cite{Murakami_PRL_2006,PhysRevB.83.121310,LiuZheng_Bi_PRL,ZhouMiao_Bi,Bi/SiC_2017}, and WTe$_2$~\cite{WTe2_Sci,WTE2_Wu_Sci} for 2D TIs (also known as quantum spin Hall insulators). However, in comparison, the proposed SOTI materials are much limited. In 3D,
predictions have been made on SnTe~\cite{Schindler2018SA}, bismuth~\cite{Schindler2018}, $X$Te$_2$ ($X=$ Mo, W)~\cite{ZJWang_WTe2}, Bi$_{2-x}$Sm$_x$Se$_3$~\cite{Yue2019ws}, EuIn$_2$As$_2$~\cite{XuYF2019}, MnBi$_{2n}$Te$_{3n+1}$~\cite{ZhangRX2020}, but the only experimental evidence so far is on bismuth~\cite{Schindler2018}. In 2D, the first example graphdiyne was predicted only recently~\cite{GDY_PRL,Lee_GDY}, and there is no experimental demonstration yet. This situation severely hinders the development of the field.
Thus, it remains a big challenge to identify realistic SOTI materials, especially for 2D.

In this work, we propose a universal approach to achieve SOTIs. The idea is to engineer an existing conventional TI by breaking the time reversal symmetry, e.g., via introducing magnetism. We show that as long as it satisfies certain conditions, the resulting system will generally become a SOTI.
The approach is especially suitable for 2D, where the magnetism can be induced by the proximity effect from a magnetic substrate.
As an example, by first-principles calculations, we show that bismuthene on EuO(111) (Bi/EuO) is a SOTI with protected corner states. Importantly, this also represents the first realistic example of a 2D magnetic SOTI. Distinct from graphdiyne which is effectively a spinless system, the topological corner states for Bi/EuO are spin polarized. Benefited from the strong spin-orbit coupling (SOC) and proximity effect, the corner states here are in a sizable boundary gap $\sim 83$ meV, greatly facilitating the experimental detection. Intriguingly, in our approach, the topological transition between first- and second-order TI phases can be controlled by the magnetic phase transition. By taking advantage of the rich inventory of the first-order TIs, our work opens a practical avenue towards realistic SOTIs with a wealth of possibilities.

{\color{blue}{\em The approach.}}---The proposed approach applies for both 3D and 2D systems. For the ease of presentation, we will focus on 2D in the following discussion. Let us begin with a conventional 2D TI, where the time reversal symmetry $(\mathcal{T})$ is preserved. To facilitate the analysis, we assume that the inversion symmetry $(\mathcal{P})$ is also preserved. In the end, we will see that the breaking of $\mathcal{P}$ is not essential to the resulting SOTI state.

A conventional TI features Dirac-type gapless boundary states. For 2D, these are the 1D spin-helical edge states, forming two spin-polarized edge bands, linearly crossing at a $\mathcal{T}$-invariant momentum (usually the $\bar{\Gamma}$ point) of the edge Brillouin zone (BZ). Around the Dirac point, the edge bands are captured by a 1D Dirac model. For example, in the Kane-Mele model (\ref{KM}) considered below, the edge is described by \begin{equation}\label{edge}
  \mathcal{H}_\text{edge}^0=vq\sigma_z,
\end{equation}
where $q$ is the momentum along the edge and the Pauli matrix $\sigma_z$ corresponds to the electron spin.

Now consider breaking the $\mathcal{T}$ symmetry in this conventional TI by introducing a ferromagnetic exchange field. We claim that as long as it remains an insulator and the exchange field is not too strong to alter the original band ordering across the gap (satisfied when $\Delta_M<\Delta$, where $\Delta_M$ is the exchange splitting and $\Delta$ is the bulk bandgap), the resulting system will generally be a SOTI.

This can be easily argued from the boundary perspective, as the Dirac point is protected by $\mathcal{T}$, it will generally be gapped by the $\mathcal{T}$-breaking perturbation (unless there exists extra crystalline symmetry protections which we need to further destroy). This gap-opening in the Dirac model is captured by a mass term $\mathcal{H}_m$. For example, for model (\ref{edge}), an exchange field along $x$ gives $\mathcal{H}_m=m\sigma_x$. It is known that the gapped Dirac model $(\mathcal{H}_\text{edge}^0+\mathcal{H}_m)$ admits a $\mathbb{Z}_2$ topological classification by the sign of its mass $m$. Therefore, a 0D corner state must appear at the intersection of two edges with masses of opposite signs, corresponding to the Jackiw-Rebbi topological domain wall mode~\cite{Jackiw_PRD}. Note that in our scheme, if one edge has a positive mass, its opposite edge related by $\mathcal{P}$ is described by
\begin{equation}
\mathcal{P}(\mathcal{H}_\text{edge}^0+\mathcal{H}_m)\mathcal{P}^{-1}=-(\mathcal{H}_\text{edge}^0-\mathcal{H}_m),
\end{equation}
i.e., it will have an effective negative mass. This shows that by our approach, the edges in the resulting system must fall into two topologically
 distinct classes, so that the topological corner states must exist.

The above analysis demonstrate that by our approach, one can transform a conventional TI into a SOTI. The discussion is general, independent of the material details, thus it is a universal approach.

Before proceeding, we have a few remarks. First, the analysis can be directly generalized to 3D, where the sample surfaces are described by 2D Dirac models, and the resulting SOTI has 1D topological hinge states at hinges between surfaces with opposite masses~\cite{Eslam2018,Piet2019,PengY2020}. Furthermore, we can also offer below an alternative argument enabled by the $\mathcal{P}$ symmetry. A 3D TI has a nontrivial (strong) $\mathbb{Z}_2$ index $\nu=1$. With $\mathcal{P}$ symmetry, $\nu$ can be determined by the Fu-Kane formula, which counts the parity of the occupied bands at the eight $\mathcal{T}$-($\mathcal{P}$-)invariant momenta $\Gamma_i$ in BZ~\cite{FuKane_PRB2007}:
\begin{equation}
  (-1)^\nu=\prod_{\Gamma_i}\prod_{\ell=1}^N\xi_{2\ell}(\Gamma_i).
\end{equation}
Because of $\mathcal{T}$, the bands form Kramers degenerate pairs at $\Gamma_i$. $\xi_{2\ell}$ is the parity eigenvalue of the $2\ell$-th band, which shares the same value $\xi_{2\ell}=\xi_{2\ell-1}$ with its Kramers partner, and $\ell$ runs through the $2N$ occupied bands below the bandgap. We notice this formula can be put into an equivalent form
\begin{equation}\label{nu}
  \nu=\sum_{\Gamma_i}\frac{n_+^{\Gamma_i}-n_-^{\Gamma_i}}{4}\qquad \text{mod}\ 2,
\end{equation}
where $n_\pm^{\Gamma_i}$ is the total number of occupied bands with positive (negative) parity eigenvalue at $\Gamma_i$. After introducing the exchange field, the Kramers degeneracies at $\Gamma_i$ are lifted, but since the perturbation does not break $\mathcal{P}$, the states at $\Gamma_i$ still retains their parity eigenvalues.
It has been shown that a 3D $\mathcal{P}$-invariant SOTI can be characterized by a $\mathbb{Z}_4$ index~\cite{PRB_Z4},
\begin{equation}\label{kappa}
  \kappa=\sum_{\Gamma_i}\frac{n_+^{\Gamma_i}-n_-^{\Gamma_i}}{2}\qquad \text{mod}\ 4,
\end{equation}
where an odd $\kappa$ indicates a metallic state, and $\kappa=2$ gives a SOTI if the bulk gap is fully opened. Since we have assumed that the exchange field does not alter the band ordering across the gap, the values of $n_\pm^{\Gamma_i}$ must remain the same.
Now, comparing Eq.~(\ref{kappa}) with Eq.~(\ref{nu}), one immediately notices that $\nu=1$ must lead to $\kappa=2$, i.e., the resulting system must be a SOTI.

Second, in practice, the  exchange field can be generated from magnetic doping or from magnetic proximity effect. The recent proposal on 3D Bi$_{2-x}$Sm$_x$Se$_3$ is actually an example of this general approach, where the magnetic doping of Sm in 3D TI Bi$_2$Se$_3$ turns the system into a SOTI~\cite{Yue2019ws}. To achieve adequate doping level while maintaining ferromagnetic ordering is a challenging task. In the following, we show that combining 2D TIs with magnetic substrates could offer a promising way to realize 2D SOTIs.

Third, we have assumed the $\mathcal{P}$ symmetry above to facilitate the analysis. Other crystal symmetries can have the same effect, if they guarantee gap inversion between different boundaries~\cite{Hughes2017,Langbehn2017,SongZD2017,Schindler2018SA,GDY_PRL}. However, it is important to note that the resulting second-boundary modes should remain robust in the presence of weak symmetry-breaking perturbations (e.g., by the substrate), as long as the bulk and edge gaps are not closed~\cite{Langbehn2017, Physics132}. This is especially clear from the edge perspective, where $\mathcal{P}$ is already broken at each edge hence will not affect the $\mathbb{Z}_2$ classification of the edge~\cite{GDY_PRL}, and to remove all the corner modes, one needs to close all edge gaps, which cannot be achieved by any weak (or local) perturbations. This point will be explicitly demonstrated in the examples below.

Finally, for experimental detection on 2D SOTIs, it is preferred to have the corner modes sitting inside the gap. This can be facilitated by the emergent chiral symmetry intrinsic to the edge Dirac model, which pins the corner modes approximately in the mid of the edge gap. Spatial symmetries of the sample can also help to ensure the degeneracy of these modes. Strong defects at the corner may push the corner modes into the bulk continuum and hinders the detection, so they should be minimized in sample preparation as much as possible.

\begin{figure}[t!]
\includegraphics[width=8.5 cm]{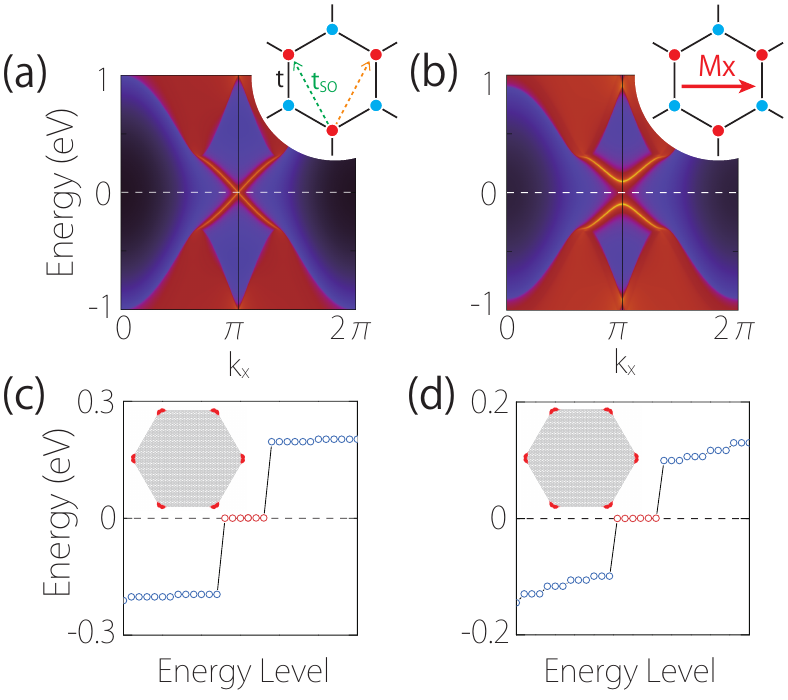}
\caption{Edge spectra for the Kane-Mele model (a) without and (b) with the exchange field ($\Delta_{M}=0.4t$). Here, $t=1$ eV and $t_\text{SO}=0.06t$. (c) Energy spectrum of a hexagonal-shaped disk for the model in (b). (d) is for the same disk geometry by further including the $\mathcal{P}$-breaking staggered potential and Rashba SOC ($\Delta_s=0.2t$ and $t_R=0.05t $). The insets in (c, d) exhibit the spatial distribution of the (red-colored) zero-modes in the spectra. }
\label{KaneMele}
\end{figure}

{\color{blue}{\em A model study.}}---We demonstrate our idea first with a paradigmatic model for 2D TIs, i.e., the Kane-Mele model~\cite{KaneMele_PRL,KaneMele_Graphene}. This model is defined on a honeycomb lattice, with
\begin{equation}\label{KM}
  H_0=\sum_{\langle ij\rangle,\alpha} tc_{i\alpha}^\dagger c_{j\alpha}+\sum_{\langle\langle ij\rangle\rangle,\alpha\beta}it_\text{SO}\nu_{ij}
\sigma^z_{\alpha\beta}c_{i\alpha}^\dagger c_{j\beta},
\end{equation}
where the first term is the nearest-neighbor hopping, and the second term is the intrinsic SOC for the second-neighbor hopping, with $\nu_{ij}=+(-)$ if the electron makes a left (right) turn during hopping from site $j$ to $i$.
This model hosts a 2D TI phase when $t_\text{SO}\neq 0$. The intrinsic SOC opens a bandgap $\Delta=6\sqrt{3}t_\text{SO}$. As shown in Fig.~\ref{KaneMele}(a), on the edge, there exists a pair of gapless spin-helical edge bands, forming a $\mathcal{T}$-protected Dirac crossing described by model (\ref{edge}).

Now we add the exchange field term
\begin{equation}
  H_M=\frac{\Delta_M}{2}\sum_i (\hat{\bm n}\cdot \bm \sigma)_{\alpha\beta}c_{i\alpha}^\dagger c_{i\beta}
\end{equation}
to the Kane-Mele model, where the unit vector $\hat{\bm n}$ denotes the direction of the exchange field. For example, considering the field in the $x$ direction, the corresponding edge spectrum is shown in Fig.~\ref{KaneMele}(b). One can see that $H_M$ opens a gap at the Dirac point, and for $\Delta_M$ small compared with $\Delta$, each edge is described by a 1D gapped Dirac model, as we have discussed.

To demonstrate the topological corner states, we need to consider a disk geometry, such as the inset in Fig.~\ref{KaneMele}(c). The calculated spectrum is also plotted. One observes that there exists six zero modes inside the boundary gap opened by $\Delta_M$. By checking their wave function distribution, one confirms that these states are localized at the six corners of the disk [see Fig.~\ref{KaneMele}(c)].

The results remain robust for any small in-plane exchange field.
It is worth noting that here, an exchange field along $z$ does not open a gap at the edge, because the model (\ref{KM}) has an extra mirror symmetry $\mathcal{M}_z=-i\sigma_z$, which offers an additional protection for the edge Dirac point. Nevertheless, this symmetry is removed by any generic exchange field with an in-plane component, or by the Rashba SOC arising from the substrate,
$ H_R=\sum_{\langle ij\rangle,\alpha\beta}it_\text{R}
(\bm\sigma\times\hat{\bm d}_{ij})^z_{\alpha\beta}c_{i\alpha}^\dagger c_{j\beta}
$, where $\hat{\bm d}_{ij}$ is the unit vector pointing from $j$ to $i$. By including $H_R$, an exchange field along $z$ will also drive the system into a SOTI.

We have argued that the resulting SOTI is robust against $\mathcal{P}$-breaking perturbations. To explicitly demonstrate this point, we consider the $\mathcal{P}$-breaking staggered potential term $H_s=(\Delta_s/2)\sum_{i,\alpha}\xi_i c_{i\alpha}^\dagger c_{i\alpha}$, with $\xi_i=\pm 1$ for the two sublattices. Figure~\ref{KaneMele}(d) shows the calculation results for the same disk geometry with $H_s$ and also $H_\text{R}$ included. It confirms that when $\Delta_s$ is small compared with $\Delta$ and $\Delta_M$, the corner states will remain robust even when $\mathcal{P}$ is broken.

\begin{figure}[t!]
\includegraphics[width=8.5 cm]{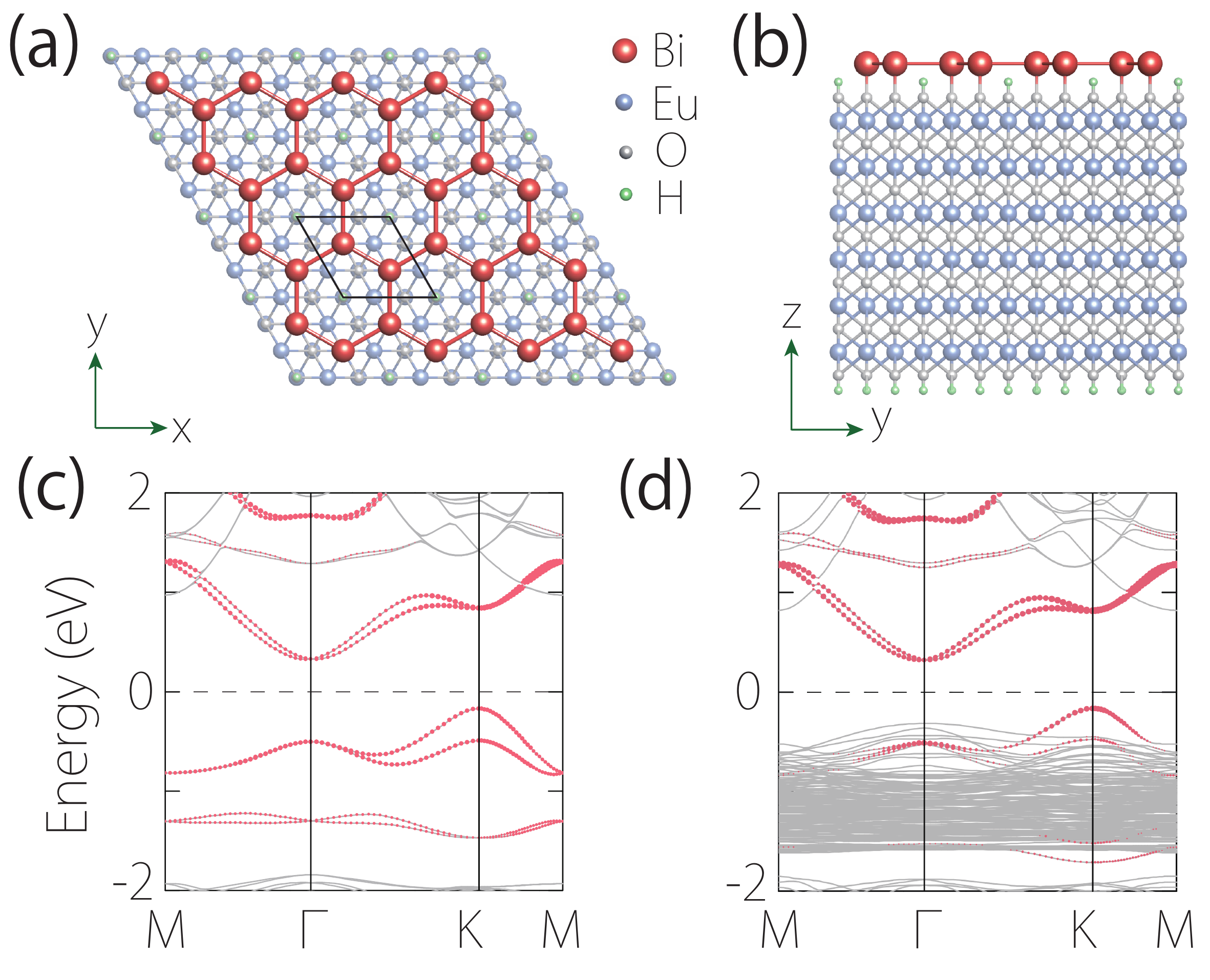}
\caption{(a) Top and (b) side view of the structure of bismuthene grown on EuO(111) substrate. (c, d) Band structure of the system (c) without and (d) with magnetism. SOC is included for both cases. The red circles indicate the weight of projection onto the Bi-$p_x$ and Bi-$p_y$ orbitals. Magnetization direction in (d) is along $x$. } \label{crystal}
\end{figure}


{\color{blue}{\em Bi/EuO.}}---Now we apply our approach to a real material system---Bi/EuO. Bismuthene refers to the 2D form of Bi. It may take several metastable crystal structures, and they are all predicted to be 2D TIs~\cite{Murakami_PRL_2006,PhysRevB.83.121310,LiuZheng_Bi_PRL,ZhouMiao_Bi,Bi/SiC_2017,LiuCC2014,Lu_npj_2016,ACSnano2020}. Recently, bismuthene was successfully realized by epitaxial growth on the SiC(0001) substrate~\cite{Bi/SiC_2017}, resulting in a planar honeycomb lattice without any buckling. The large gap $\sim0.8$ eV and the topological edge states have been confirmed in experiment.

\begin{figure}[t!]
\includegraphics[width=8.5 cm]{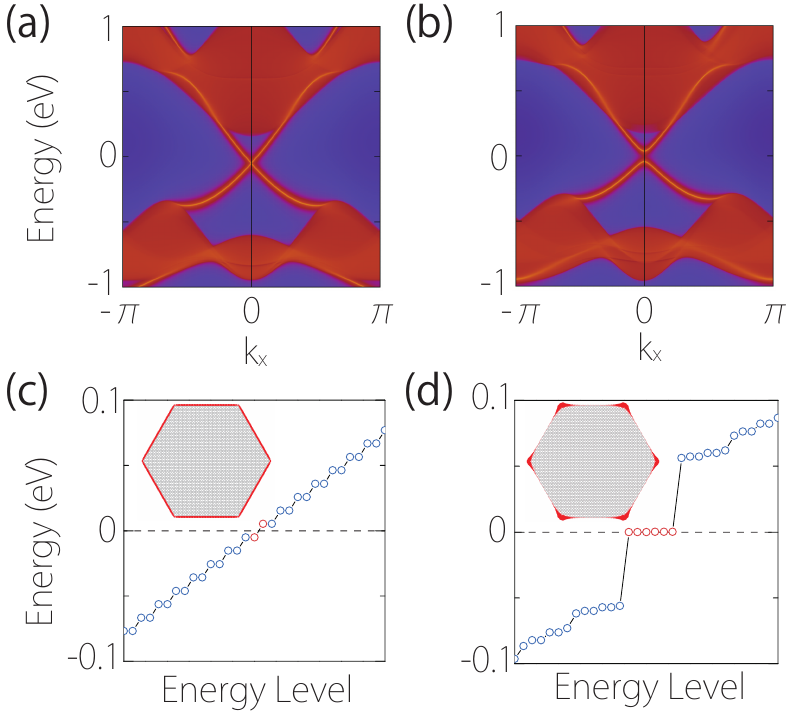}
\caption{Edge spectra of Bi/EuO (a) without and (b) with magnetism. (c, d) Spectra for the nanodisk geometry (c) without and (d) with magnetism.
The insets show the distribution of the states marked in red color in the spectra. }
\label{edges}
\end{figure}

Given its large TI gap $\Delta$, according to our theory, the SOTI state can be readily achieved in bismuthene by introducing a moderate exchange field. Here, we consider case by proximity coupling to the magnetic substrate EuO. EuO is a large-gap ferromagnetic insulator with Curie temperature $T_C\sim 69$ K~\cite{VanVleck1961,McGuire1964,EuO}. In thin film form, its magnetization prefers the in-plane direction~\cite{EuO,Mairoser2015vh}. We consider bismuthene grown on the EuO(111) surface which shares a similar surface geometry as SiC(0001). Our first-principles calculation based on the density functional theory (DFT) (see Supplemental Material for details~\cite{SM}) shows that the resulting bismuthene layer also forms a planar honeycomb lattice [with ($\sqrt{3}\times \sqrt{3}$ R30$^{\circ}$) superstructure, see Fig.~\ref{crystal}(a, b)], similar to the case with SiC substrate~\cite{Bi/SiC_2017}.

Figure~\ref{crystal}(c) shows the calculated band structure when the magnetism in the substrate is turned off, which simulates the situation for $T>T_C$. One finds that the system is an insulator with a {bandgap $\sim$ 0.5 eV}. The low-energy bands are mainly from the $p_x$ and $p_y$ orbitals of Bi layer. And an evaluation of the $\mathbb{Z}_2$ invariant confirms that the system is a 2D TI.

Next, we turn on the magnetism in EuO. The result is shown in Fig.~\ref{crystal}(d). One observes that the band edges are still dominated by Bi and are only slightly altered. The estimated exchange splitting is about $0.15$ eV, which is comparable to that found in WS$_2$/EuS systems (0.2 to 0.3 eV)~\cite{Norden:2019wh}. In the calculation, we can gradually increase the magnetic moment from zero to its equilibrium value. By monitoring the band structure change during this process, we confirm that there is no inversion of band ordering across the gap. Therefore, according to our theory, the ground state of Bi/EuO must be a magnetic SOTI.

To verify the edge spectrum and the corner states, we construct an \emph{ab initio} tight-binding model for the low-energy bands based on the DFT result and the Wannier functions~\cite{SM}, which fully includes the substrate effects. Figures~\ref{edges}(a, b) show a comparison of the edge spectra without and with magnetism, from which one observes that the proximity-induced magnetic exchange opens a sizable boundary gap $\sim 83$ meV for the originally gapless edge bands.  In Figs.~\ref{edges}(c, d), we show the results for the nanodisk geometry. Without magnetism, the spectrum is quasi-continuous (the double degeneracy is due to $\mathcal{T}$) and the states at the Fermi level correspond to the quantum spin Hall edge states [Fig.~\ref{edges}(c)]. With magnetism, six zero modes clearly emerges at the Fermi level inside the boundary gap. The wave function distribution plotted in Fig.~\ref{edges}(d) confirms that they are the topological corner states.

\begin{figure}[t!]
\includegraphics[width=8.5 cm]{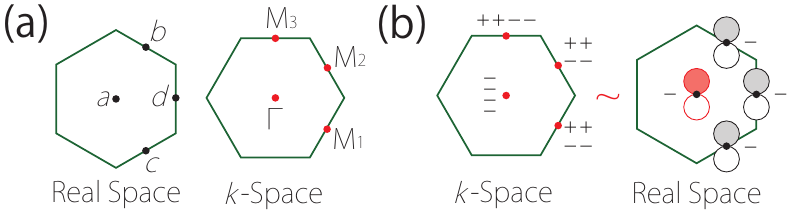}
\caption{(a) Left: Four $\mathcal{P}$-invariant Wyckoff sites in real space. Right: Four $\mathcal{P}$-invariant momenta in $k$-space. 
(b) Left panel shows the parity eigenvalues of the occupied low-energy Bi bands. The bands are equivalent to four odd local symmetric orbitals, one on each site, as schematically shown in the right panel.}
\label{EBR}
\end{figure}

{\color{blue}{\em Discussion.}}---We have proposed a universal approach to realize SOTIs by using conventional TIs.
Distinct from most previous studies, the SOTI here is with $\mathcal{T}$ breaking. Particularly, the proposed Bi/EuO represents the first
realistic 2D magnetic SOTI. This feature leads to distinct physical properties. For example, in Bi/EuO, the corner states are spin polarized; whereas in the previous example graphdiyne, they are not~\cite{GDY_PRL}. Furthermore, it is intriguing that the magnetic and the topological phase transitions here are tied together: the magnetic phase transition is simultaneously a transition between first-order TI and SOTI.

The corner states in Bi/EuO can also be understood from an alternative perspective of topological quantum chemistry (TQC)~\cite{Bradlyn2017wf,Kruthoff2017,Po2017ts,Song_RSI}.  Because, as we mentioned, the substrate does not alter the band ordering across the gap in Bi, one can simplify the analysis by first neglecting the $\mathcal{P}$-breaking effect from the substrate but retaining the exchange field (which breaks most crystal symmetries except for $\mathcal{P}$). In real space, we hence consider the four $\mathcal{P}$-invariant Wyckoff sites, labeled as $a, b, c, d$ in a unit cell [Fig.~\ref{EBR}(a)]. Local orbitals at such site are either even or odd under $\mathcal{P}$, known as symmetric orbitals~\cite{SongZD2017}. TQC maps the symmetry representations of such orbitals to the band structure in $k$-space.
Reversely, one can extract the local orbital properties by decomposing a band representation using the elementary ones. Here, the four occupied Bi-$p$ bands have parity eigenvalues as shown in Fig.~\ref{EBR}(b). Analysis with TQC shows that, in terms of parity, the system is equivalent to that of four odd symmetric orbitals at the $a, b, c, d$ sites [Fig.~\ref{EBR}(b)]. Notice that the $a$ site is at the center of the hexagon, not occupied by an atom nor a bond, and the corner states we obtained in Fig.~\ref{edges}(d) correspond to the fractional filling of this $a$ orbital, known as the filling anomaly in the previous study~\cite{wieder2018axion}. Again, once created, the corner states will remain robust against $\mathcal{P}$-breaking as long as such perturbations do not close the boundary gap.

Experimentally, we expect that Bi/EuO can be fabricated by using similar epitaxial growth method as in Ref.~\cite{Bi/SiC_2017}. The topological corner states can be probed by the scanning tunneling spectroscopy (STS), which maps the local density of states (at liquid-helium temperature, STS can reach a resolution of 1 meV). The spin polarization of the corner states can also be detected by using a ferromagnetic tip in STS. By tuning the substrate magnetism, e.g., by varying $T$ across $T_C$, one can observe the transition between the 0D corner states and the 1D spin-helical edge states.

\begin{acknowledgments}
\textit{Acknowledgments---} The authors thank D. L. Deng for helpful discussions. This work is supported by the NSFC (Grants No. 11834014, No.11774018, No. 11504013), and the Singapore MOE AcRF Tier 2 (MOE2019-T2-1-001).
\end{acknowledgments}	

\bibliographystyle{apsrev4-1}{}
\bibliography{SOTI_ref}



\end{document}